\documentclass[twocolumn]{aastex62}

\received{January 1, 2018}
\revised{January 7, 2018}
\accepted{\today}
\submitjournal{ApJ}

\shorttitle{V4046 Sgr}
\shortauthors{ Ru\'iz-Rodr\'iguez et al.}

\begin{document}

\title{Constraints on a Putative Planet Sculpting the V4046 Sagittarii Circumbinary Disk}

\correspondingauthor{D. Ru\'iz-Rodr\'iguez}
\email{darpci@rit.edu}

\author[0000-0002-0786-7307]{Dary Ru\'iz-Rodr\'iguez}
\affil{Chester F. Carlson Center for Imaging Science, School of Physics $\&$ Astronomy, and Laboratory for Multiwavelength Astrophysics,\\
Rochester Institute of Technology, 54 Lomb Memorial Drive, Rochester NY 14623 USA.}

\author{Joel H. Kastner}
\affiliation{Chester F. Carlson Center for Imaging Science, School of Physics $\&$ Astronomy, and Laboratory for Multiwavelength Astrophysics,\\
Rochester Institute of Technology, 54 Lomb Memorial Drive, Rochester NY 14623 USA.}


\author{Ruobing Dong}
\affiliation{Department of Physics $\&$ Astronomy, University of Victoria Victoria BC V8P 1A1 Canada.}

\author{David A. Principe}
\affiliation{Kavli Institute for Astrophysics and Space Research, Massachusetts Institute of Technology, Cambridge, MA 02139, USA.}

\author{Sean M. Andrews}
\affiliation{Harvard-Smithsonian Center for Astrophysics, 60 Garden Street, Cambridge, MA, 02138, USA.}

\author{David J. Wilner}
\affiliation{Harvard-Smithsonian Center for Astrophysics, 60 Garden Street, Cambridge, MA, 02138, USA.}




\begin{abstract}

We analyze the highest-resolution millimeter continuum and near-infrared (NIR) scattered-light images presented to date of the circumbinary disk orbiting V4046 Sgr, a $\sim$20 Myr old actively accreting, close binary T Tauri star system located a mere 72.4 pc from Earth. We observed the disk with the Atacama Large Millimeter/submillimeter Array (ALMA) at 870 $\micron$ during Cycle 4, and we analyze these data in conjunction with archival NIR (H band) polarimetric images obtained with SPHERE/IRDIS on the ESO Very Large Telescope. At 0.3$''$ (20 au) resolution, the 870 $\micron$ image reveals a marginally resolved ring that peaks at $\sim$32 au and has an extension of $\sim$ 90 au. We infer a lower limit on dust mass of $\sim$ 60.0 M$_{\oplus}$ within the 870 $\micron$ ring, and confirm that the ring is well aligned with the larger-scale gaseous disk. A second, inner dust ring is also tentatively detected in the ALMA observations; its position appears coincident with the inner ($\sim$14 au radius) ring detected in scattered light. Using synthetic 870 $\micron$ and H-band images obtained from disk-planet interaction simulations, we attempt to constrain the mass of the putative planet orbiting at 20 au. Our trials suggest that a circumbinary Jovian-mass planet may be responsible for generating the dust ring and gap structures detected within the disk. We discuss the longevity of the gas-rich disk orbiting V4046 Sgr in the context of the binary nature of the system.

\end{abstract}

\keywords{Millimeter Obsevations, Scattered light, Circumbinary Disks, Planetary Formation}


\section{Introduction} \label{sec:intro}

Observations of circumstellar disks composed of cold dust and molecular gas around young ($<$ 30 Myr) stellar objects provide crucial information about the formation of planets. Ideally, direct detections of newborn planets within such disks may provide fundamental constraints on planet formation theories. However, current observational tools are limited by a range of complicating factors such as contrast ratios and inner working angles, and, as a result, indirect detection methods are needed to predict when, where and how planets form. Observations with high spatial resolution can detect signposts of a forming planet, such as disk gaps \citep[e.g][]{Andrews2016} or spiral arms \citep[e.g][]{Benisty2015} resulting from planet-disk gravitational interactions. Comparing such structures with theoretical models of disk-planet interactions can provide essential parametrizations and characterizations of the formation and evolution of planetary systems \citep{Fung2015, Dong2017}.

Furthermore, most Sun-like stars form in binary or multiple systems
\citep{Duquennoy1991}, some of which will host circumbinary
disks. The study of such disks is necessary to determine whether
their conditions are conducive to the formation of circumbinary
planets. Indeed, the lifetimes of circumbinary
disks may exceed those of disks orbiting single stars
\citep{Alexander2012}. The time available for the formation of a
circumbinary planet, and its location of formation, should indicate
the likely building mechanism, i.e., core accretion \citep{Pollack1996} vs. gravitational instability \citep{Boss1997}.

The circumbinary, protoplanetary disk orbiting the nearby, actively
accreting pre-main sequence binary system V4046 Sgr \citep[][and
references therein]{Kastner2018} offers the highly unusual opportunity
to explore the observational signatures of circumbinary planet-disk
interactions. The V4046 Sgr system lies at a mere 72.47$\pm$0.34 pc
\citep{Gaia2018}, and is a member of the $\beta$ Pictoris moving group
\citep{Zuckerman2004}, fixing its age at 23$\pm$3 Myr
\citep{Mamajek2014}. The central binary consists of a nearly
equal-mass pair of K-type stars with masses of $\sim$0.90
$\pm$ 0.05 $\rm M_{\odot}$  and 0.85 $\pm$ 0.04 $\rm M_{\odot}$
\citep{Rosenfeld2012} in a tight and nearly circular orbit \citep[P $\sim$
2.4 d, $e$$\lesssim$ 0.01;][]{StempelsGaum2004}. \citet{Rodriguez2010} and  \citet{Rosenfeld2013} used Submillimeter Array (SMA) interferometry to establish that the V4046 Sgr binary system is surrounded
by a massive ($\sim$0.1 $M_\odot$), gas-rich circumbinary disk
extending to $\sim$300 au. As delineated in the
subarcsecond ALMA molecular line imaging study by \citet{Kastner2018},
the molecular disk is characterized by extended, centrally peaked CO
and HCN emission and a sequence of sharp and diffuse rings of emission
from HC$_{3}$N, CH$_{3}$CN, DCN, DCO$^{+}$, C$_{2}$H, N$_{2}$H$^{+}$, and H$_{2}$CO.

From analysis of SMA 1.3 mm continuum observations, \citet{Rosenfeld2013} reported a large inner hole of
$\sim$ 30 au with the majority of the dust mass residing in a narrow ring centered at 37 au. Subsequent ALMA mm-wave imaging at $\sim$0.5$''$ resolution has confirmed this result  \citep{Huang2017, Guzman2017, Bergner2018, Kastner2018}. Coronagraphic/polarimetric $\sim$ 3 au resolution Gemini
Planet Imager (GPI) imaging of scattered light from dust grains
revealed the presence of a double-ring structure in the dust
distribution \citep{Rapson2015}. The double-ring morphology presented
an inner cavity $\sim$10 au in radius, a narrow ring with a peak flux
at $\sim$14 au and a dust gap at $\sim$ 20 au, where the second ring
begins and extends to 45 au. More recently, confirming the double-ring
scattered light structure, \citet{Avenhaus2018} reported rings
centered at $\sim$15 au and $\sim$27 au, on the basis of SPHERE/IRDIS polarimetric differential imaging in the J and H bands.

Near Infrared (NIR) polarized and millimeter observations serve as complementary probes of the surface structure and distribution of material in the midplane regions of a disk, allowing the identification of the ``dust filtration'' effect resulting from the pressure maximum outside a gap opened by a forming planet \citep[]{Pinilla2012, Dong2012}. Although considerable effort has been invested in generating high-fidelity simulations of NIR scattered-light and mm continuum observations \citep[e.g.][]{Debes2013, Dong2018}, examples of attempts to model real disks in both wavelength regimes simultaneously remain few and far between \citep[e.g.][]{Baruteau2019}.

In this paper, we present the results of simultaneous comparisons of SPHERE and ALMA data with detailed two-dimensional two fluid (gas + particle) hydrodynamical calculations coupled with three-dimensional Monte Carlo Radiative Transfer simulations \citep{Dong2012, Dong2015}, so as to explore the observational signatures of gaps possibly opened by a single planet in the V4046 Sgr circumbinary disk. This paper is organized as follows. In Section \ref{Sec:observations}, we describe the observations and the data reduction process, while in Section \ref{Sec:results}, we present the main results. In Section \ref{Sec:Morphology}, we describe the morphology and main features of V4046 Sgr detected by ALMA and SPHERE data. In Section \ref{Sec:Modelling}, the disk model and Monte Carlo Radiative Transfer (MCRT) simulations are presented, followed by a discussion in Section \ref{Sec:Discussion}. Section \ref{Sec:Summary} summarizes our main results and conclusions.

\section{OBSERVATIONS AND DATA REDUCTION} 
\label{Sec:observations}

\begin{figure*}
     \begin{center}
%
       
           { \label{Fig:Almadata} 
           \textbf{\Large \qquad   870 \micron     \qquad  \qquad \qquad    \qquad  1.65 \micron   \qquad }\par
            \includegraphics[width=0.37\textwidth]{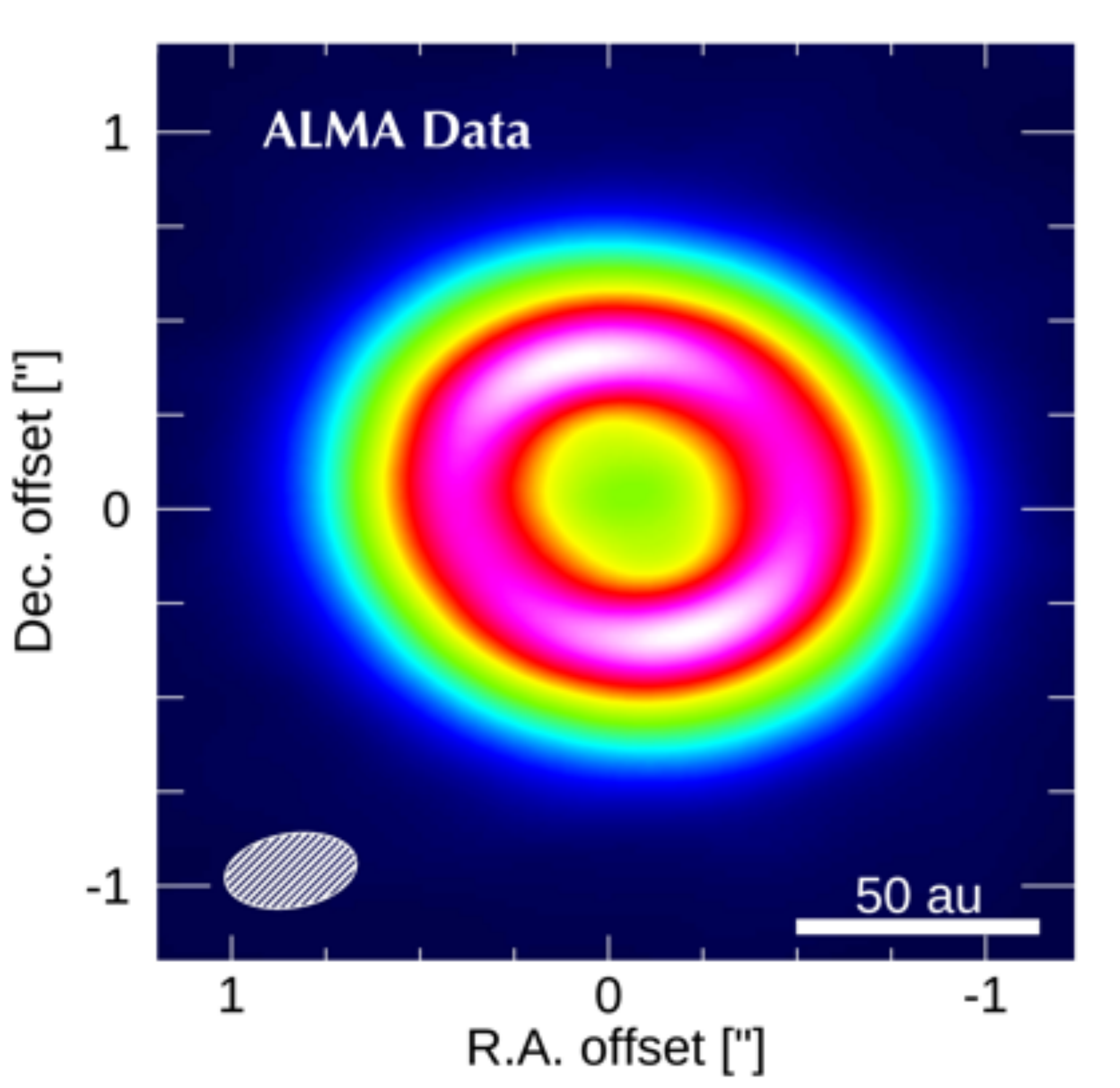}
        }%
          { \label{fig:SPHEREdata}
           \includegraphics[width=0.365\textwidth]{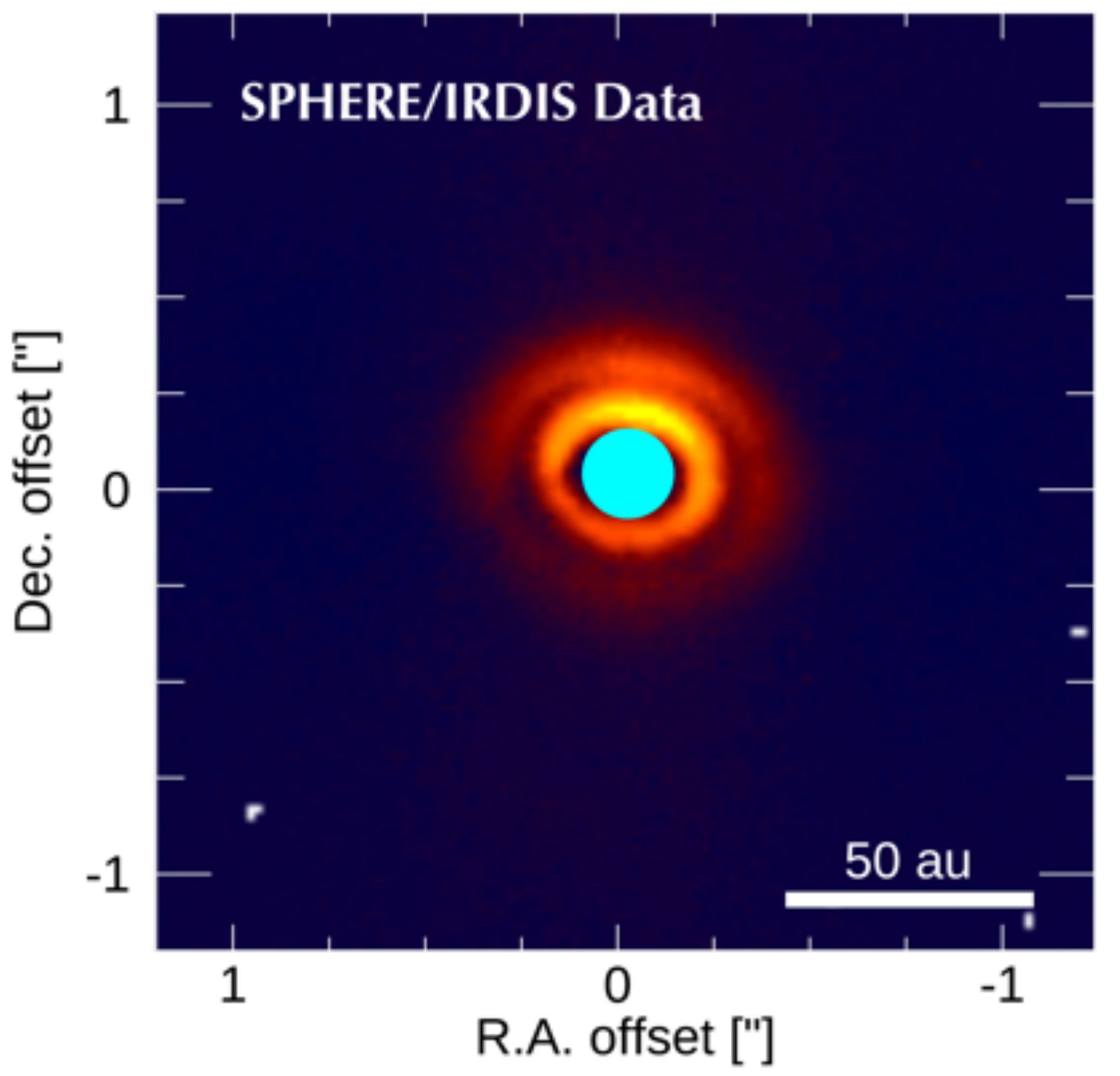}
        }%
          {  \label{fig:ALMAmodel}
            \includegraphics[width=0.37\textwidth]{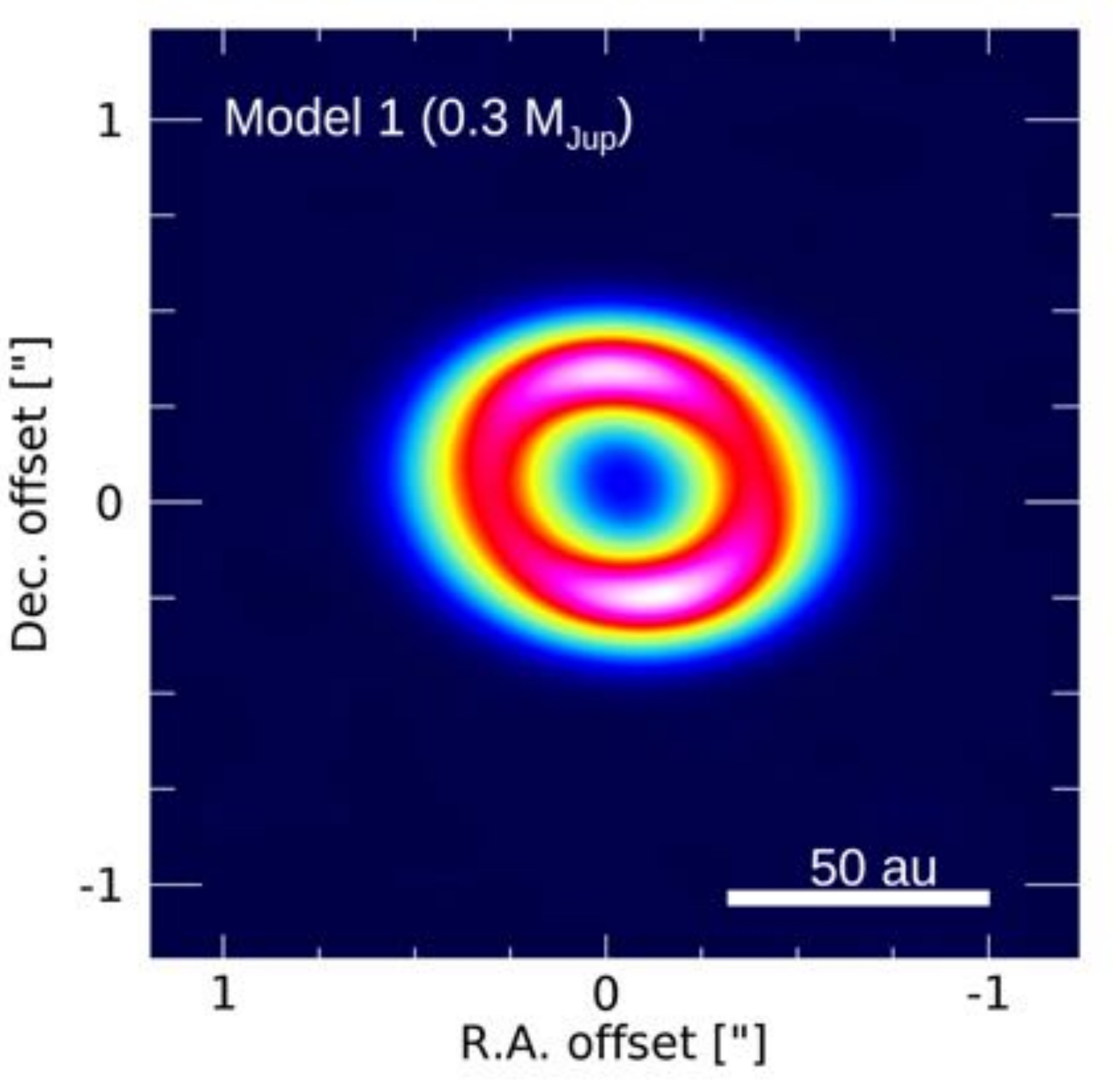}
        }
          { \label{fig:SPHEREmodel}
            \includegraphics[width=0.365\textwidth]{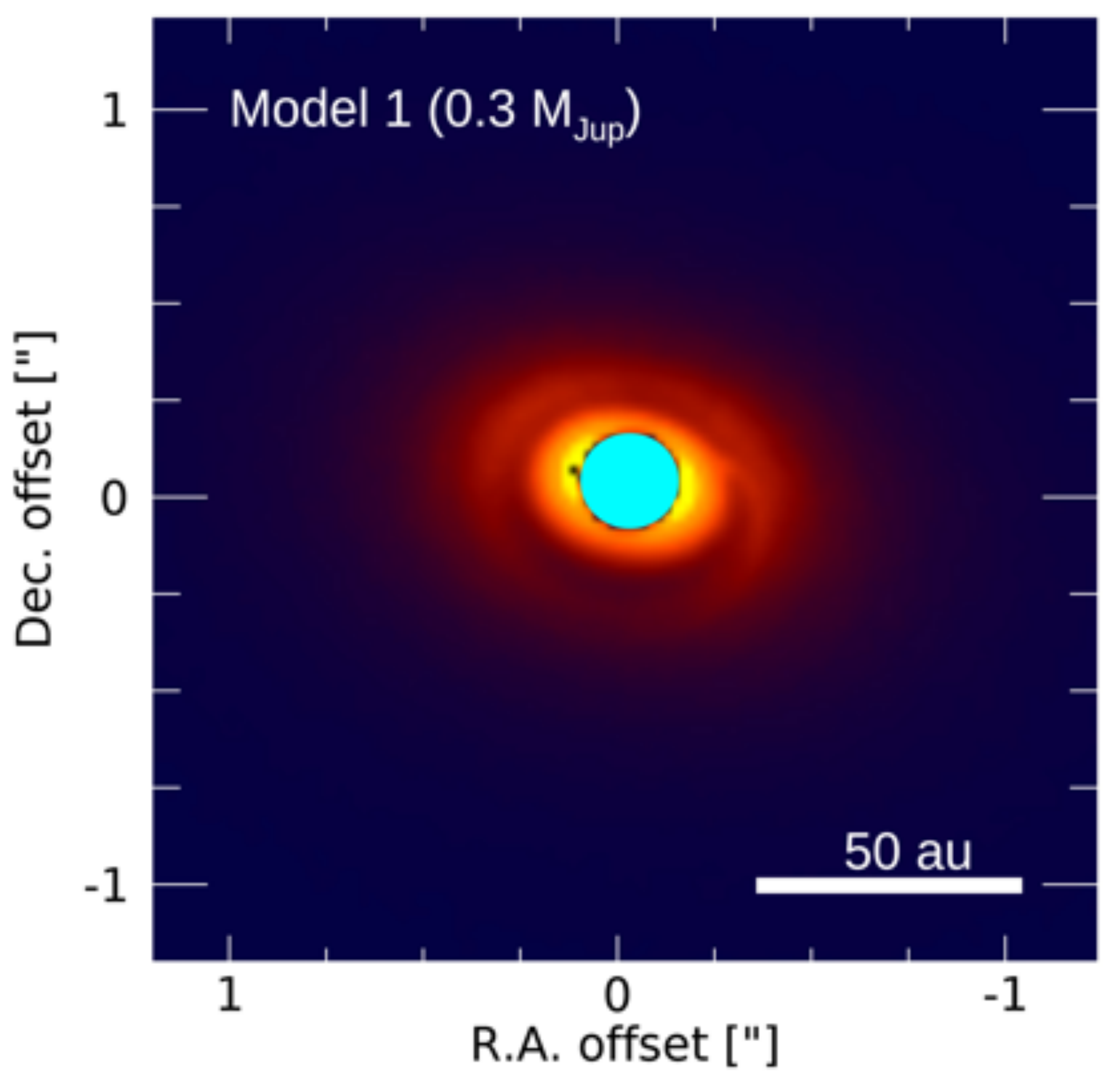}
         }%
           {  \label{fig:ALMAmodel}
            \includegraphics[width=0.37\textwidth]{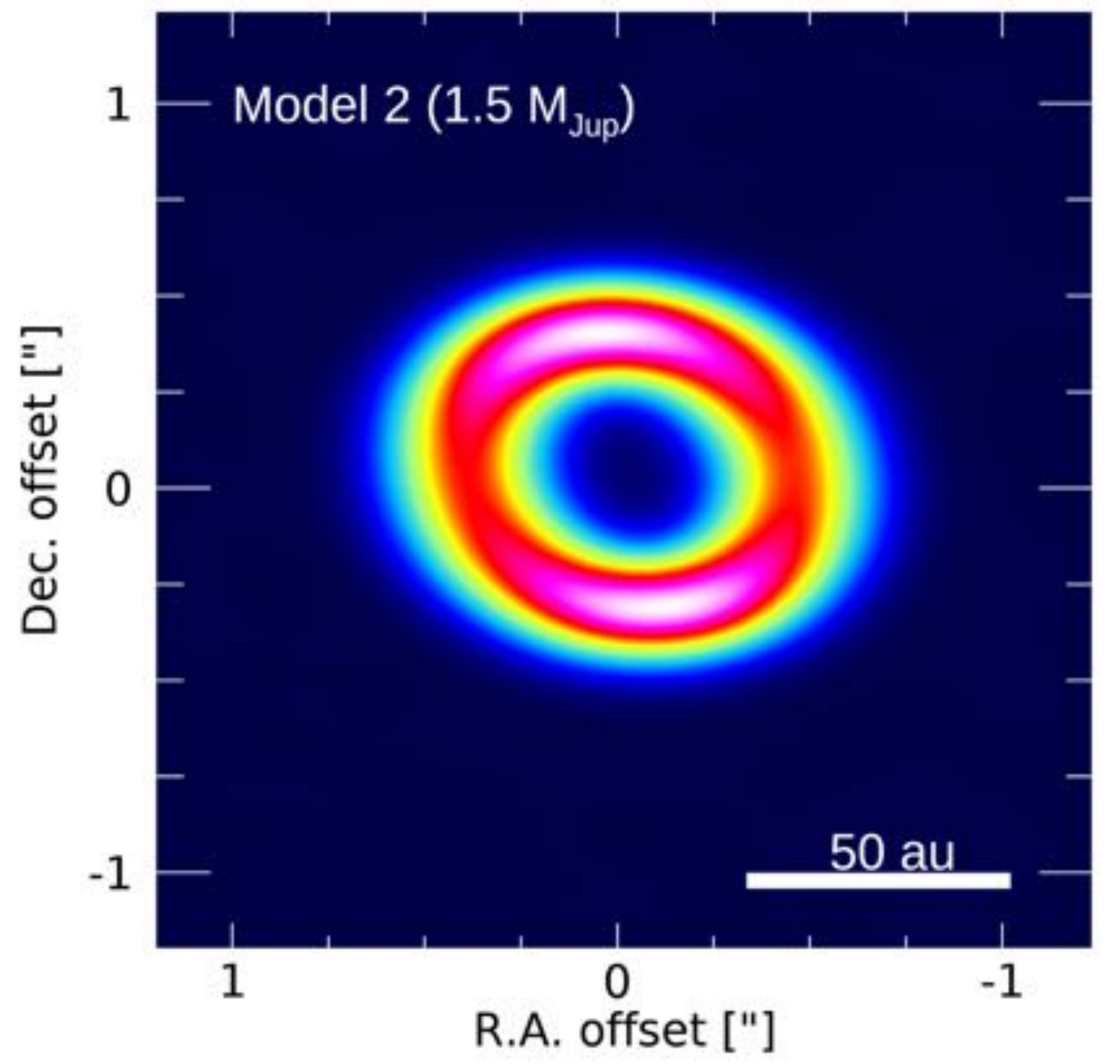}
        }
          { \label{fig:SPHEREmodel}
            \includegraphics[width=0.37\textwidth]{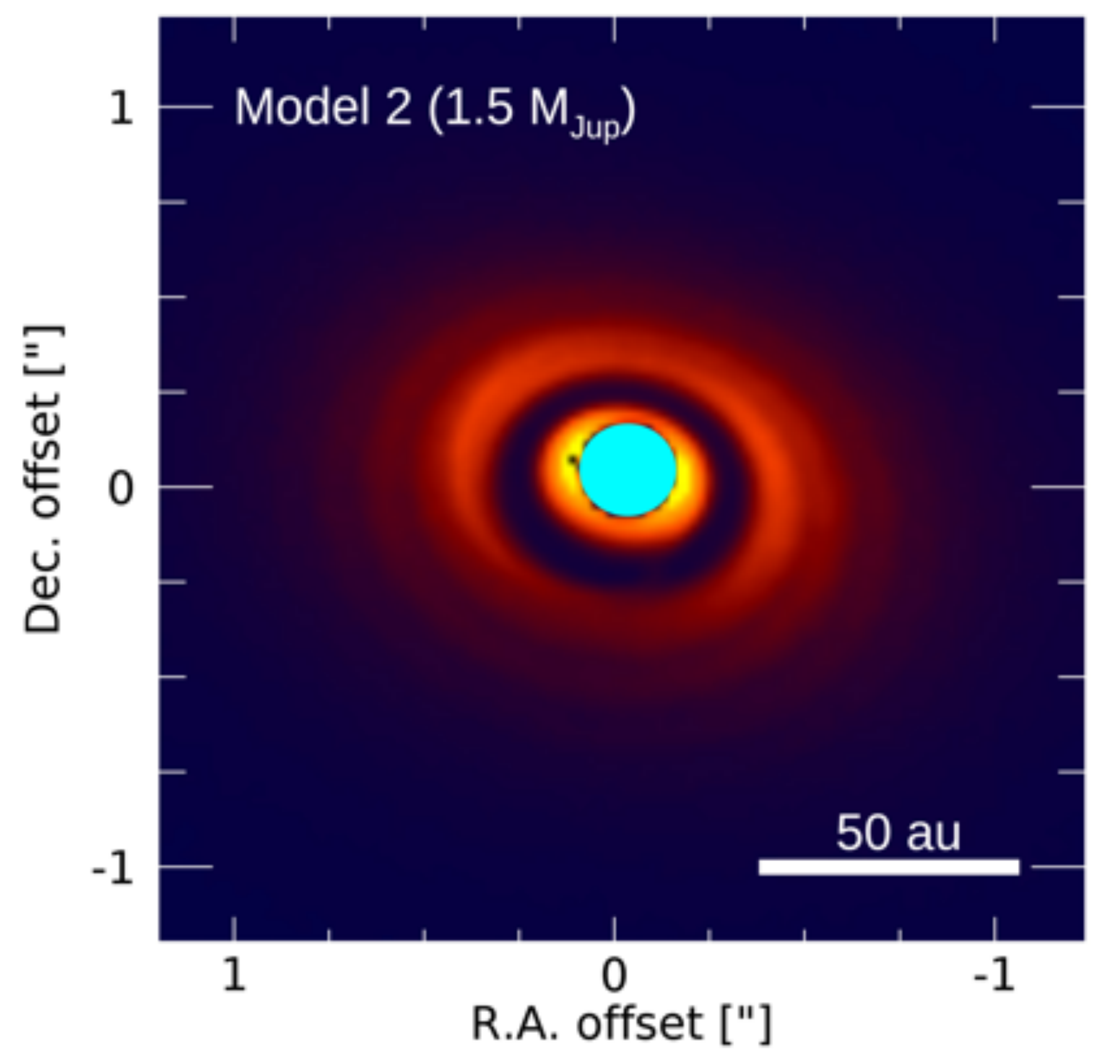}
         }%
         \vspace{-10pt}
    \end{center}
    \caption{Comparison of simulated images and observations at 345
      GHz and 1.65 $\micron$ of the circumbinary disk orbiting V4046
      Sgr. From top to bottom: ALMA Band 7 and SPHERE/IRDIS
      obsevations; convolved MCRT simulation using a planet mass of
      0.3 M$_{\rm J}$ (Model 1); convolved MCRT simulation using a
      planet mass of 1.5 M$_{\rm J}$ (Model 2). Top left panel:  870
      $\micron$ continuum image of V4046 Sgr. The white dashed ellipse
      in the lower left-hand corner indicates the size of the
      synthesized beam (0.29$^{''}$$\times$ 0.17$^{''}$ at
      P.A. -79.7$^{\rm o}$). Top right panel: SPHERE/IRDIS H-Band
      image with a blue filled circle representing the N ALC YJH S
      coronagraph (inner working angle $\sim$ 0.1$^{''}$ = 7.3 au at
      72.4 pc ). The synthetic ALMA Band 7 (left) and SPHERE H-band (right) images in the middle and bottom rows were generated by convolving the MCRT simulations with a Gaussian point-spread function, producing convolved images with angular resolution of $\sim$ 0.2$^{''}$ and $\sim$ 0.04$^{''}$, respectively. Convolved images are scaled so that the planet is at 20 au. The blue circle in the two synthetic H-band images indicates the size and position of the artificial coronagraph used in the simulations. North is up and east is to the left in all images. The color scale is linear.}
   \label{Fig:SPHEREALMA}
\end{figure*}

\subsection{ALMA Observations}
\label{Sec:ALMAObservations}

ALMA Band 7 (345 GHz) observations of V4046 Sgr were obtained on May
09, 2017 as part of the Cycle 4 program 2016.1.00315.S. The phase
center of the observations is $\alpha$ (J2000) = 18$^{\rm h}$ 14$^{\rm
  m}$ 10.47$^{\rm s}$ ; $\delta$ (J2000) = -32$^{\rm o}$47$^{\rm '}$
34.50$^{\rm ''}$. The array configuration was C40-5 with a
  longest baseline of 1.1 km. The correlator was set up with four
spectral windows in dual polarization mode, centered at 330.616 GHz,
345.825 GHz, 344.031 GHz, and 332.531GHz, and the bandwidths used were
937.5, 468.8, 2000.0 and 2000.0 MHz, respectively. The total on-source integration time was 11.7 minutes.  Analysis of the CO isotopologue emission covered by the first two (narrower) basebands at 330 and 345 GHz will be presented in a forthcoming paper (Ruiz-Rodriguez et al., in preparation). In this paper, we
focus exclusively on the data obtained in the latter two (continuum mode) configurations.

The broadband (2 GHz baseband) 332 and 344 GHz visibility data
analyzed here were edited, calibrated and imaged using the pipeline
version r39732 in CASA 4.7.2. The quasar J1826-2924  was observed as
phase calibrator, J1823-3454 was used as flux calibrator, while the
quasar J1924-2914 was observed for bandpass calibration. We
  applied self-calibration with three rounds of phase calibration, and
  used the TCLEAN algorithm to image the data using two different
  Briggs weighting values. First, we set the Briggs weighting (robust)
parameter $R$ to
$-0.5$, to achieve a balance between resolution and sensitivity. The
resulting rms was 0.10 mJy beam$^{-1}$ within a bandwidth of 5.29 GHz
and a synthesized beam of 0.29$^{''}$$\times$ 0.17$^{''}$  at
P.A. -79.7$^{\rm o}$. Second, we applied uniform weighting ($R = -2$)
to optimize the spatial resolution, resulting in a synthesized beam of 0.2$^{''}$ $\times$ 0.1$^{''}$  at P.A. $\sim$-79.7$^{\rm o}$ with rms of 0.18 mJy beam$^{-1}$. The images were constructed on a 256 $\times$ 256 pixel grid with 20 mas pixel size. The two Briggs weighting values provide comparable integrated flux densities. For most of the analysis presented here, we consider the image reconstruction using $R = -0.5$, to prioritize signal-to-noise over spatial resolution. The $R = -2.0$ image reconstruction is analyzed in Section \ref{Sec:Briggs2}.

\subsection{SPHERE Observations}

The archival H-band polarimetric images of V4046 Sgr presented in this work were obtained on March 13, 2016 with the ESO Very Large Telescope (VLT) SPHERE-IRDIS instrument using the N ALC YJH S coronagraph.  Images were taken with the BB$\_$H filter in polarimetric differential imaging (PDI) mode with a total integration time of 3072 seconds. A complete description of these data was presented in \citet{Avenhaus2018}.  

The SPHERE data were reduced and analyzed using the EsoReflex pipeline (v. 2.8.5) and the SPHERE IRDIS workflow (v. 0.31.0) where data were collected, organized, and reduced to account for darks, flats, star centering and de-rotation.  Polarimetric images were then further reduced following the procedure described in \citet{Avenhaus2014, Avenhaus2018}. The SPHERE instrument separates the beam into two orthogonal states, the so-called ordinary and extraordinary beams. A pre-correction for instrumental polarization was performed by normalizing the flux in the ordinary and extraordinary beams using the presumed unpolarized halo emission signature from the central star. We determine the halo ordinary and extraordinary flux ratio $X_{o/e}$ = (f$_{o}$ / f$_{e}$) in an aperture with inner and outer radius of 47 and 72 pixels from the image center, respectively, and then multiply the extraordinary beam by ($X _{o/e}$)$^{1/2}$ and the ordinary beam by ($X _{o/e})$$^{-1/2}$.


We calculate the Stokes vectors following  \citet{Avenhaus2014, Avenhaus2018}:
\begin{equation} 
{p_q = \frac{R_Q - 1}{R_Q + 1}, \hspace{4mm} p_u = \frac{R_U - 1}{R_U + 1}}
\end{equation}
with
\begin{equation}
\small 
{R_Q = \sqrt{ \frac{  I_{\textrm{ord}}^{0^{\circ}} / I_{\textrm{extra}}^{0^{\circ}}   }{  I_{\textrm{ord}}^{-45^{\circ}} / I_{\textrm{extra}}^{-45^{\circ}} }  }, \hspace{0mm}  R_U = \sqrt{ \frac{  I_{\textrm{ord}}^{-22.5^{\circ}} / I_{\textrm{extra}}^{-22.5^{\circ}}   }{  I_{\textrm{ord}}^{-67.5^{\circ}} / I_{\textrm{extra}}^{-67.5^{\circ}} }  } }
\end{equation}
 where the subscripts indicate the ordinary and extraordinary beams while the superscripts indicate the position of the half-wave plate angle.  The Stokes Q and U are then determined by:
\begin{equation}
{Q = p_q \times I_Q, \hspace{4mm} U = p_u \times I_U}
\end{equation}
where the total intensity (I) is:

\begin{equation}
\small
{I_Q = (I_{\textrm{ord}}^{0^{\circ}} +  I_{\textrm{extra}}^{0^{\circ}}   +  I_{\textrm{ord}}^{-45^{\circ}}   +  I_{\textrm{extra}}^{-45^{\circ}}     ) / 2}
\end{equation}
\begin{equation}
\small
{I_U = (I_{\textrm{ord}}^{-22.5^{\circ}} +  I_{\textrm{extra}}^{-22.5^{\circ}}   +  I_{\textrm{ord}}^{-67.5^{\circ}}   +  I_{\textrm{extra}}^{-67.5^{\circ}}     ) / 2}.
\end{equation}

Assuming single scattering events, the light scattered from the disk should be linearly polarized in the azimuthal direction so we use the radial stokes parameters $Q_{\phi}$ and $U_{\phi}$:
\begin{equation}
{Q_\phi = +Q \textrm{cos}(2\phi) + U \textrm{sin}(2\phi)}
\end{equation}
\begin{equation}
{U_\phi = -Q \textrm{sin}(2\phi) + U \textrm{cos}(2\phi) }
\end{equation}
\begin{equation}
{\phi = \textrm{arctan}(\frac{x - x_0}{y - y_0}) + \gamma}
\end{equation}
where $\phi$ is the angle between up on the detector and a line from the star (at position $x_0$, $y_0$) to a position on the detector.  The $\gamma$ offset angle optimizes the reduction to correct for potential misalignment of the of the half-wave plate or rotated polarization.

\section{RESULTS} 
\label{Sec:results}

In this section, we present and analyze the ALMA and SPHERE data
individually, to characterize the radial and azimuthal structure of
the disk in continuum emission and in scattered light (respectively). 
First, we describe and quantify the ALMA 870 $\micron$ continuum data
in terms of their main parameters (i.e. emission morphology, flux density,
minimum dust mass; Sec.\ 3.1). 
Then we parameterize radial profiles
extracted from the SPHERE H band data so as to estimate the sizes and locations of the gaps and
rings observed in scattered light (Sec.\ 3.2). 

\subsection{Continuum Emission at 870 $\micron$}
\label{Sec:Continuum}

The ALMA Cycle 4 Band 7 image of the intermediate inclination V4046
Sgr circumbinary disk is displayed in the top left panel of Figure
\ref{Fig:SPHEREALMA}. The 870 $\micron$ continuum image reveals a
well-defined ring with a large central hole. We measure the radius and
width of this ring in two stages: first, obtain estimates of the
inclination and the position angle (P.A.) of the ring, by fitting a
surface brightness model in the visibility domain (Sec.\ 3.1.1); second, we
use these parameters to deproject the ring and obtain its radius and
width from Gaussian fitting of the resulting image plane radial profile (Sec.\ 3.1.2).

\subsubsection{Inclination and P.A.}
\label{Sec:IncPA}

Considering that the dust continuum emission at 870 $\micron$
is resolved, and it is concentrated into a ring that shows an azimuthal uniformity in intensity, we estimated the inclination, position angle and continuum flux density by fitting an elliptical Gaussian directly to the visibility data. To that end, we
used the CASA routine \textit{uvmodelfit} \citep{Marti2014}, which fits a single
component source model  (i.e., point-source, Gaussian, or
disk) to the ($u,v$) visibility data. 

This fit yields a disk inclination of 32.42$^{\circ}$ $\pm$ 0.07 and a P.A. of 74.33$^{\circ}$$\pm$ 0.14. We verified the inclination by using only short baselines within the $u,v$ range 0--450 k$\lambda$ (location of the null in the real part of the visibilities). In addition, we find an 870 $\micron$ continuum flux density of 876.30 $\pm$ 0.51 mJy, where the uncertainty is the formal error resulting from the fit.

\subsubsection{Radial Profiles}
\label{Sec:RadialProfile}

We used this best-fit inclination and P.A. (32.42$^{\circ}$ $\pm$ 0.07
and 74.33$^{\circ}$ $\pm$ 0.14, respectively) to deproject the ALMA and
SPHERE images. We then extracted radial profiles from the deprojected
images and performed Gaussian fitting, so as to characterize the properties of the flux maxima. 

\begin{figure*}
     \begin{center}
%
           { \label{fig:first}
            \includegraphics[width=0.49\textwidth]{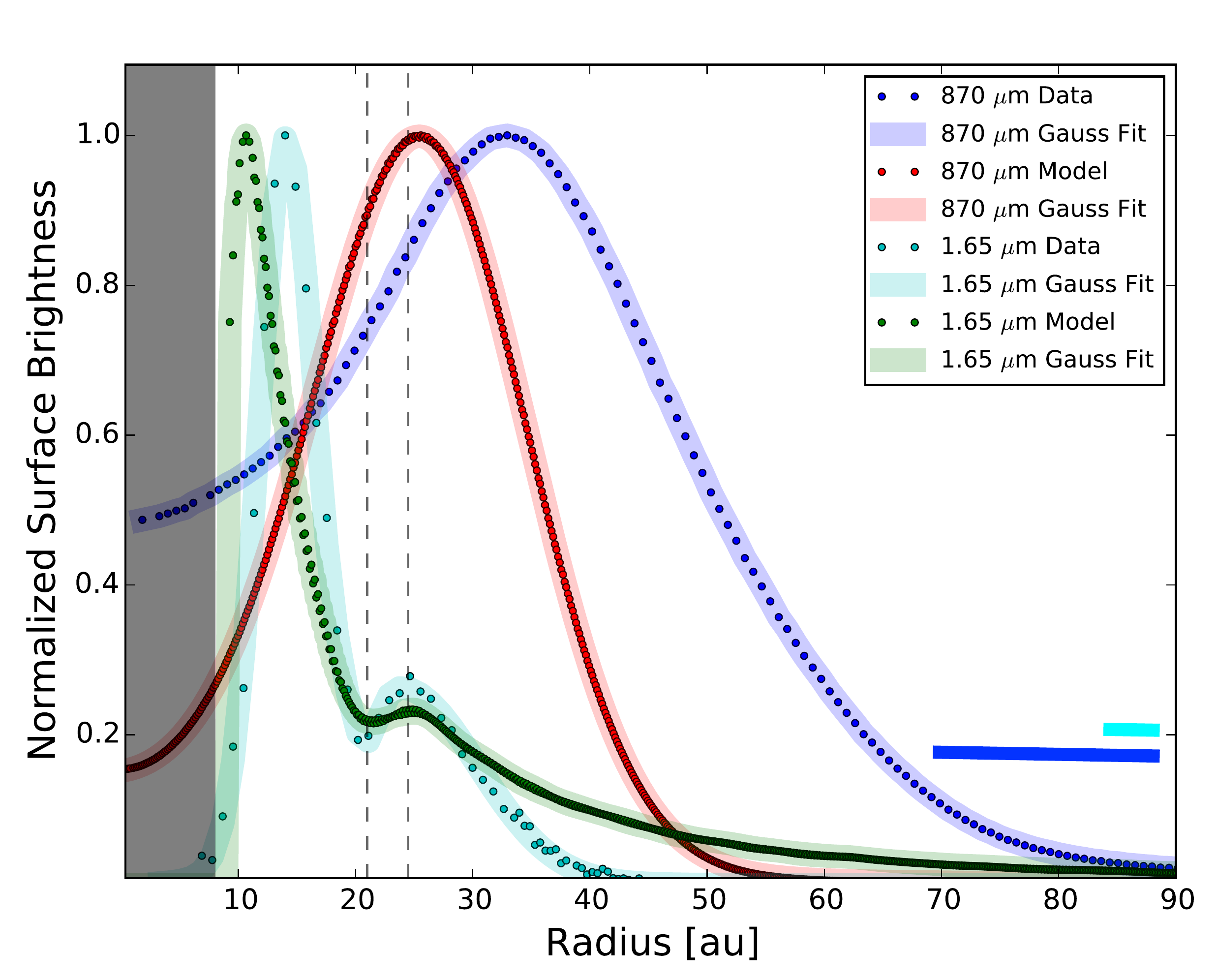}
        }%
          { \label{fig:second}
           \includegraphics[width=0.49\textwidth]{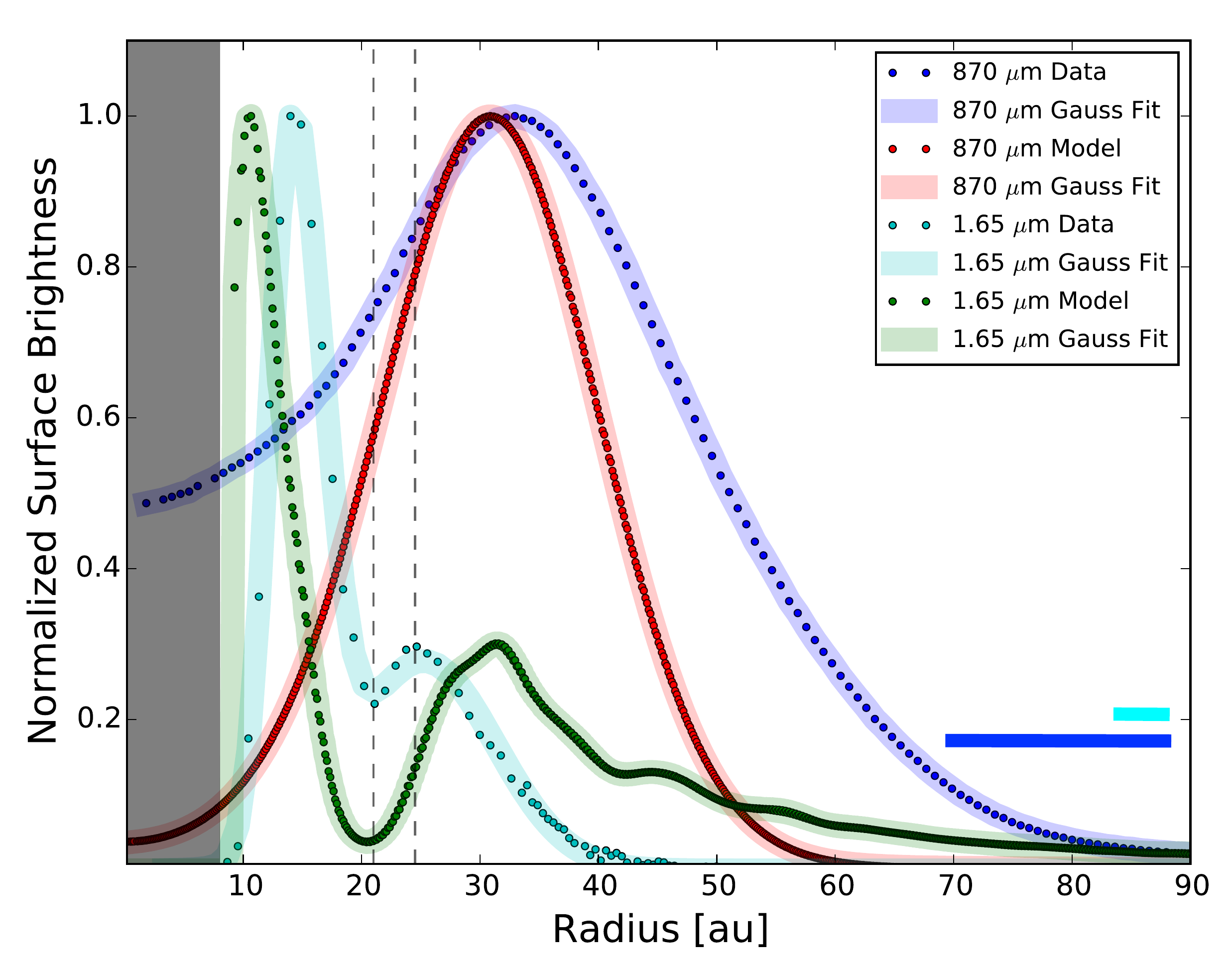}
        }%
    \end{center}
    \caption{  
    Comparison of the surface brightness profiles extracted from the
    deprojected synthetic and observed 
    H-band and 870 $\micron$ continuum images (see Figure
    \ref{Fig:SPHEREALMA} and Section \ref{Sec:RadialProfile}). The left and right
    panels show comparisons for models with
    planet masses of 0.3 M$_{\rm J}$ and 1.5 M$_{\rm J}$
    respectively (Models 1 and 2; Sec.\ 5.2). The dotted curves indicate
    the radial profiles extracted from the data and models, while the colored shading indicates the
    Gaussian fits to these profiles, with color-coding as in the
    Figure legends (i.e., cyan and blue shading for the observed SPHERE H-band
    scattered light and ALMA 870 $\micron$ continuum intensity
    profiles, and green and red shading for their respective Model 1 and 2 counterparts).
    The grey shaded area represents the
    inner working angle of the coronagraph used in
    the SPHERE/IRDIS observations (i.e., $\sim$ 0.1$^{''}$, or 7.3 au
    at 72.4 pc) . The vertical dashed lines indicate
    the location of the gap (intensity minimum) at $\sim$20 au and the gap outer edge at
    $\sim$25 au. The blue and cyan bars in the lower right-hand corner of each frame indicate the spatial resolution at each wavelength regime, i.e., $\sim$ 20 au for the ALMA data and $\sim$ 4 au for the SPHERE data, respectively.}
  
   \label{Fig:Radialprofile}
\end{figure*}

The radial profiles from the ALMA continuum images were azimuthally
averaged, while the radial profiles from the SPHERE scattered-light images were
obtained as the averages over wedges with opening
angles of 20$^{\circ}$ oriented along the disk major axis (see Section \ref{Sec:SPHEREdata}). The resulting radial profiles and Gaussian fits are displayed in
Figure \ref{Fig:Radialprofile}. We achieved these fits with a superposition of an average of three Gaussians for every profile, where this number of components is required to account for the significant asymmetries in the profiles. The best-fit parameters are listed in
Table \ref{Table:Radialprofile}. We find that, for the $R=-0.5$ beam size of 20
au$\times$12 au (at 72.4 pc), the 870 $\micron$ continuum emission
peaks at 32.30 $\pm$ 0.14 au, with a FWHM of 37.30 $\pm$ 2.74 au and
outer radius of $\sim$ 90 au. 

The comparison of ALMA and SPHERE
continuum emission features and morphologies is described in more detail in Section
\ref{Sec:Morphology}.

\begin{deluxetable*}{CCCCCCCCCCC}
\tabletypesize{\scriptsize}
\tablecolumns{11}
\tablenum{1}
\tablewidth{0pt}
\tablecaption{Observed and synthetic ALMA and SPHERE images: ring feature parameters\tablenotemark{a} \label{Table:Radialprofile}}
\tablehead{
\colhead{Parameter} &
\multicolumn3c{ALMA}&&\multicolumn6c{SPHERE}\\
\cline{2-4}
\cline{6-11}
\colhead{} &
\colhead{Data} &\multicolumn2c{Model}&&\multicolumn2c{Data} &\multicolumn4c{Model}\\
\cline{3-4}
\cline{6-7}
\cline{8-11}
&&\colhead{0.3 $\rm M_{Jup.}$} &\colhead{1.5 $\rm M_{Jup.}$} &&& &\multicolumn2c{0.3 $\rm M_{Jup.}$} &\multicolumn2c{1.5 $\rm M_{Jup.}$} \\
\cline{8-11}
&&&&&\colhead{Ring 1 }&\colhead{Ring 2}&\colhead{Ring 1 }&\colhead{Ring 2}&\colhead{Ring 1 }&\colhead{Ring 2}
}
\startdata
\rm Radius \  [au]  & 32.30 $\pm$ 0.14 & 26.62 $\pm$ 0.01 & 31.90 $\pm$ 0.03 && 14.10 $\pm$ 0.01 & 24.62 $\pm$ 0.08 & 10.90 $\pm$ 0.12 & 24.67 $\pm$ 0.20 & 10.62 $\pm$ 0.02 & 29.33 $\pm$ 0.19  \\
\rm FWHM  \  [au]& 37.30 $\pm$ 2.74 & 23.94 $\pm$ 0.01 & 24.43 $\pm$ 0.03 & & 6.65 $\pm$ 0.08 & 3.70 $\pm$ 0.01 & 6.26 $\pm$ 0.60 & 4.75 $\pm$ 0.09&5.13 $\pm$ 0.43 & 11.20 $\pm$ 0.33 \\
\enddata
\tablenotetext{a}{Radial locations and widths of image ring features, as obtained from Gaussian parameterizations of the surface brightness profiles of deprojected 870 $\micron$ continuum (ALMA) and H-band scattered-light (SPHERE) synthetic and observed images (see Figure \ref{Fig:Radialprofile} and Section \ref{Sec:RadialProfile})}
\end{deluxetable*}

\subsubsection{Disk Mass}
\label{Sec:Diskmass}

In the optically thin emission regime, dust mass can be estimated from the measurement of the (sub-)millimeter continuum flux at a given frequency $F_{\nu}$ \citep{Hildebrand1983}. Specifically, assuming isothermal emission and the canonical interstellar medium (ISM) gas-to-dust ratio of $\sim$100 \citep{Bohlin1978}, the dust mass (M$_{dust}$) can be related to the integrated 870  $\micron$ flux, F$_{870  \micron}$, via

\begin{equation}
{M_{dust}  =  \frac{F_\nu d^2}{\kappa_\nu B_\nu (T_{dust})} \approx 70\left \{ \frac{F_{\nu}(870)}{\rm Jy} \right \}\rm M_{\oplus }},
\label{eq:dust}
\end{equation}
where $d$ is the distance to the source, $\kappa_{\nu}$ is the dust grain opacity (we adopt 0.02 $\rm cm^{2} g^{-1}$ at 870 $\micron$ with $\beta$ = 1.5; \citet{Beckwith1990}) and $B_\nu (T_{dust}$) is the Planck function at a characteristic dust temperature (T$_{dust}$).

To estimate $T_{dust}$, we adopt the brightness temperature
calculated from the peak flux per beam, $T_B \sim 15$ K. 
From Eq.~\ref{eq:dust}, we thereby obtain a dust mass of $M_{dust}
\sim 60$ $\rm M_{\oplus }$ for the dust ring imaged by
ALMA. If we instead adopt the expected
equilibrium temperature expected for a dust ring located
$\sim$30 au from the V4046 Sgr binary, $T_{dust} \sim 30$ K, the
dust mass inferred from Eq.~\ref{eq:dust} would be a factor 2
smaller. However, the fact that the peak observed
brightness temperature $T_B$ is within a factor two of the
estimated equilibrium dust temperature indicates that the 870 $\mu$m
emission is likely optically thick. Hence, the
estimate $M_{dust}
\sim 60$ $\rm M_{\oplus }$ obtained from Eq.~\ref{eq:dust} assuming
$T_{dust} = 15$ K most likely represents a lower
limit on the dust mass within the ring imaged by ALMA.

\subsection{Scattered Light at 1.65 $\micron$}
\label{Sec:SPHEREdata}

The top right panel of Figure \ref{Fig:SPHEREALMA} shows SPHERE imaging of the V4046 Sgr circumbinary disk, revealing two rings, a cavity and a gap in scattered light \citep{Rapson2015, Avenhaus2018}. In the SPHERE images, as in previous GPI imaging \citep{Rapson2015}, the surface brightness of the rings show an asymmetry between the northern and southern sides. This surface brightness asymmetry is most likely due to preferential forward scattering by the dust grains along the line-of-sight \citep[e.g.][]{Schneider2009}. Hence, we confirm the conclusion of \citet{Rapson2015} that the observed brightness asymmetry is an indication that the disk is tipped such that the northern side is closer to Earth.

Because the scattered-light asymmetry is confined to the minor axis of the (projected) disk, we averaged over wedges with opening angle of 20$^{\circ}$ along the disk major axis to generate radial profiles of the surface brightness \citep[e.g.][]{Dong2017}. From the parametrization of the radial profiles (Section \ref{Sec:RadialProfile}), we obtain an inner cavity radius of $\sim$ 9.8 au. This is well outside the inner working angle of the coronagraph ($\sim$ 0.1'' = 7.3 au at 72.4 pc). In addition, we determine that ring 1 (inner ring) and 2 (gap outer edge) are located at 14.10 $\pm$ 0.01 and 24.62 $\pm$ 0.08 au, respectively, from the central binary (Table \ref{Table:Radialprofile}), with an inter-ring gap width of $\sim$5 au. The ring peak locations determined by  \citet{Avenhaus2018} are somewhat larger than determined here, because those authors scaled the surface brightness by r$^{2}$ to remove the effects of a diluted stellar radiation field, and they analyzed azimuthally averaged surface brightness curves.

\section{Morphology}

\label{Sec:Morphology}

\subsection{ALMA vs. SPHERE Images: Main Features}
\label{Sec:Combination}

\begin{figure}
     \begin{center}
%
           { \label{fig:first}
            \includegraphics[width=0.48\textwidth]{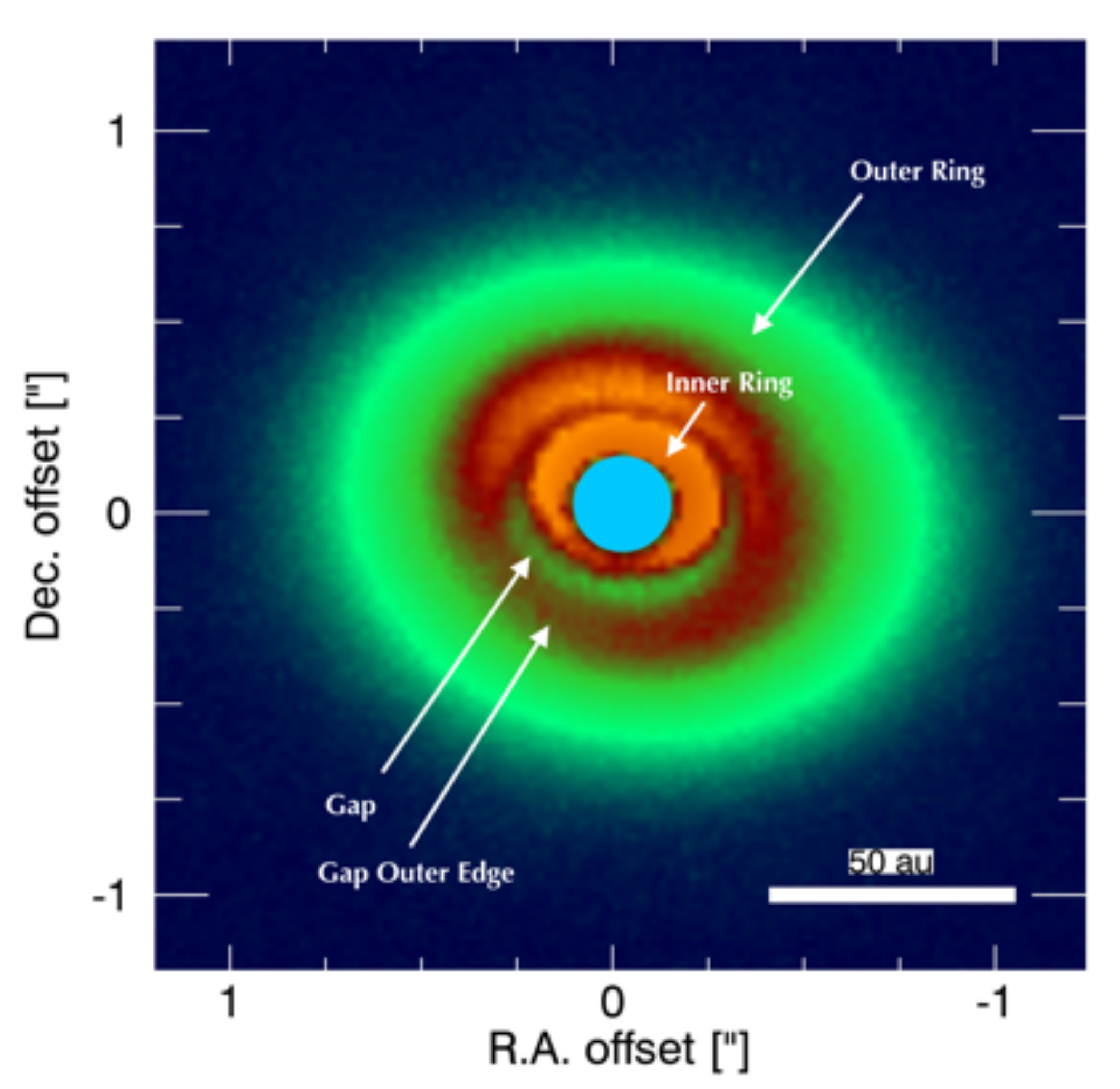}
        }%
    \end{center}
    \caption{%
      Color overlay of ALMA 870 $\micron$  continuum (green) and
      SPHERE/IRDIS 1.65 $\micron$ (red) images of the V4046 Sgr disk,  annotated with key image
      features discussed in the text. 
The blue circle indicates the image region that is hidden by the coronagraph used in the SPHERE observations.
     }%
   \label{Fig:RGB}
\end{figure}

An overlay of  the ALMA and SPHERE images for V4046 Sgr is shown in
Figure \ref{Fig:RGB}. The latter (NIR scattered light) observations trace
micron-sized particles residing at the disk surface high above the
midplane, while the former (millimeter continuum) observations are more sensitive to larger
particles ($>$100 $\micron$) near the midplane. Recent studies
comparing scattered light and millimeter continuum data have revealed
a wavelength dependency in the location of cavity and gap edges,
wherein millimeter continuum observations trace larger cavities and
gap edges than scattered light \citep[e.g.][]{Uyama2018,
  Hendler2018}. This is the result of the close coupling between gas
and micron-sized dust particles and the dearth of mm-sized dust within
the cavities and gaps. 

From the surface brightness profiles (see Figure
\ref{Fig:Radialprofile} and Table \ref{Table:Radialprofile}), we deduce that the inner scattering wall of
the ring traced by the 870 $\micron$ emission is located at a radius
of $\sim$25 au from the central binary, while the intensity
minimum between inner and outer rings is located at $\sim$20 au. We
draw vertical lines in Figure \ref{Fig:Radialprofile} indicating the
location of this gap and the outer gap edge, which corresponds to the inner scattering wall of
the 870 $\micron$ continuum emission ring (see annotations in Figure \ref{Fig:RGB}).

In theory, the dust distribution near the disk surface follows a power law
profile described by $ \frac{h}{r}$ $\sim$ $r^{\beta}$, where the
disk scale height is modulated by a flaring index $\beta$ $>$ 0,
an irradiated disk would have a typical value for the flaring
parameter of $\sim$ 1.3 \citep{Chiang1997}. In their analysis of the SPHERE scattered light imaging, \citet{Avenhaus2018} obtained estimates of flaring indexes of $\sim$0.1 for ring 1 and $\sim$1.6 for ring
2. The latter is similar to the theoretical value, indicating a flared
disk. This is surprising given the fact that at an age $\sim$ 20 Myr,
one expects significant settling of material to the midplane. However,
it is important to consider that the theoretical flaring index of 1.3
depends on temperature and assumes a uniform dust-to-gas ratio
throughout the disk.

If the micron-sized grain population is
well coupled to the gas, then there may exist a considerable small particle population
in the disk of V4046 Sgr that potentially reaches to the outer edge
of the molecular disk.

This is the case for TW Hya, where scattered
light observations trace a significant population of small dust
particles 

out to at least  230 au \citep{vanBoekel2017}. Indeed, it is possible that the relatively large
flaring index of 1.6 found for the outer scattered-light ring of the V4046 Sgr disk indicates that the
particles near the gap outer edge are well dispersed from
the disk plane, where they can scatter starlight efficiently,
while at radii larger than 60 au, the disk may become undetectable in
scattered light due to self-shadowing and not necessarily because the
disk lacks small grains at these radii.   If self-shadowing
explains the lack of surface brightness beyond 60 au, we would expect
the disk surface to lie in a more expanded and flattened shape and, as noted,
the small-grain disk may extend to, or even beyond, the gas disk
detected in CO \citep[i.e., to $\sim$300 au;][]{Kastner2018}.

\subsection{An Inner Ring in the mm Continuum?}
\label{Sec:Briggs2}

\begin{figure}
     \begin{center}
%
           { \label{fig:first}
            \includegraphics[width=0.48\textwidth]{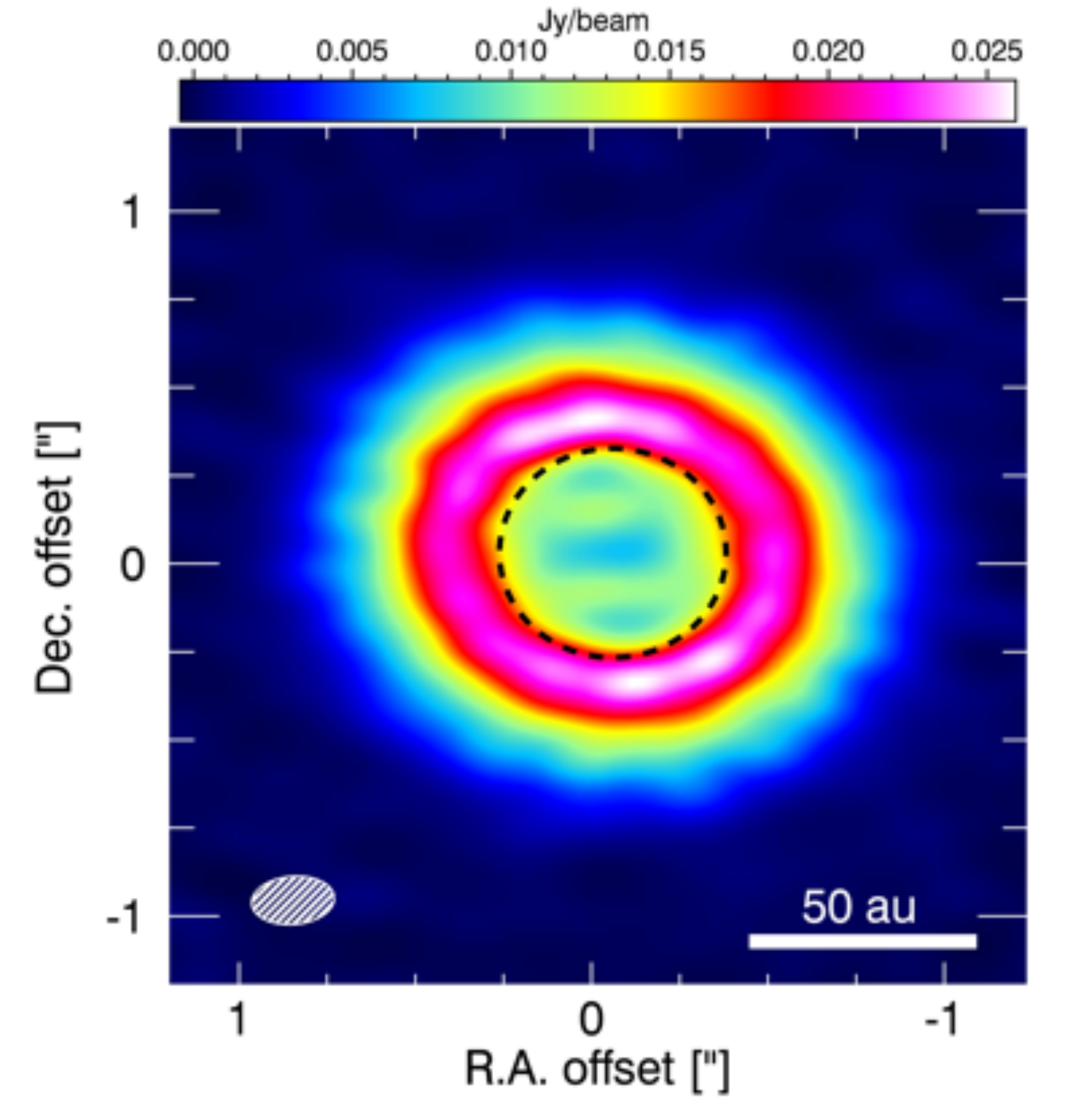}
        }%
    \end{center}
    \caption{%
       ALMA 870 $\micron$ continuum image of the V4046 Sgr disk obtained using a Briggs parameter of $-$2.0. The black dashed ellipse indicates the aperture used for the interior region integrated flux-density. The white dashed ellipse in the lower left-hand corner indicates the size of the synthesized beam (0.2$^{''}$X 0.1$^{''}$ at P.A. $\sim$-79.7$^{\rm o}$). North is up and east is to the left.
     }%
   \label{Fig:ALMASPHERE002}
\end{figure}

In Figure \ref{Fig:ALMASPHERE002}, we display the ALMA 870 $\micron$ continuum image generated with a Briggs parameter of $-2.0$, to optimize spatial resolution rather than sensitivity. From this image, we obtain an integrated flux over the elliptical region
interior to the main ring (dashed ellipse in Figure
\ref{Fig:ALMASPHERE002}) of 83.0 $\pm$ 8.0 mJy (assuming a 10\% flux
calibration accuracy). This integrated flux can be ascribed to the
marginally resolved structure(s) revealed by the ALMA millimeter continuum data
within $\sim$20 au of the central binary. 

\begin{figure}
     \begin{center}
%
          { \label{fig:second}
           \includegraphics[width=0.49\textwidth]{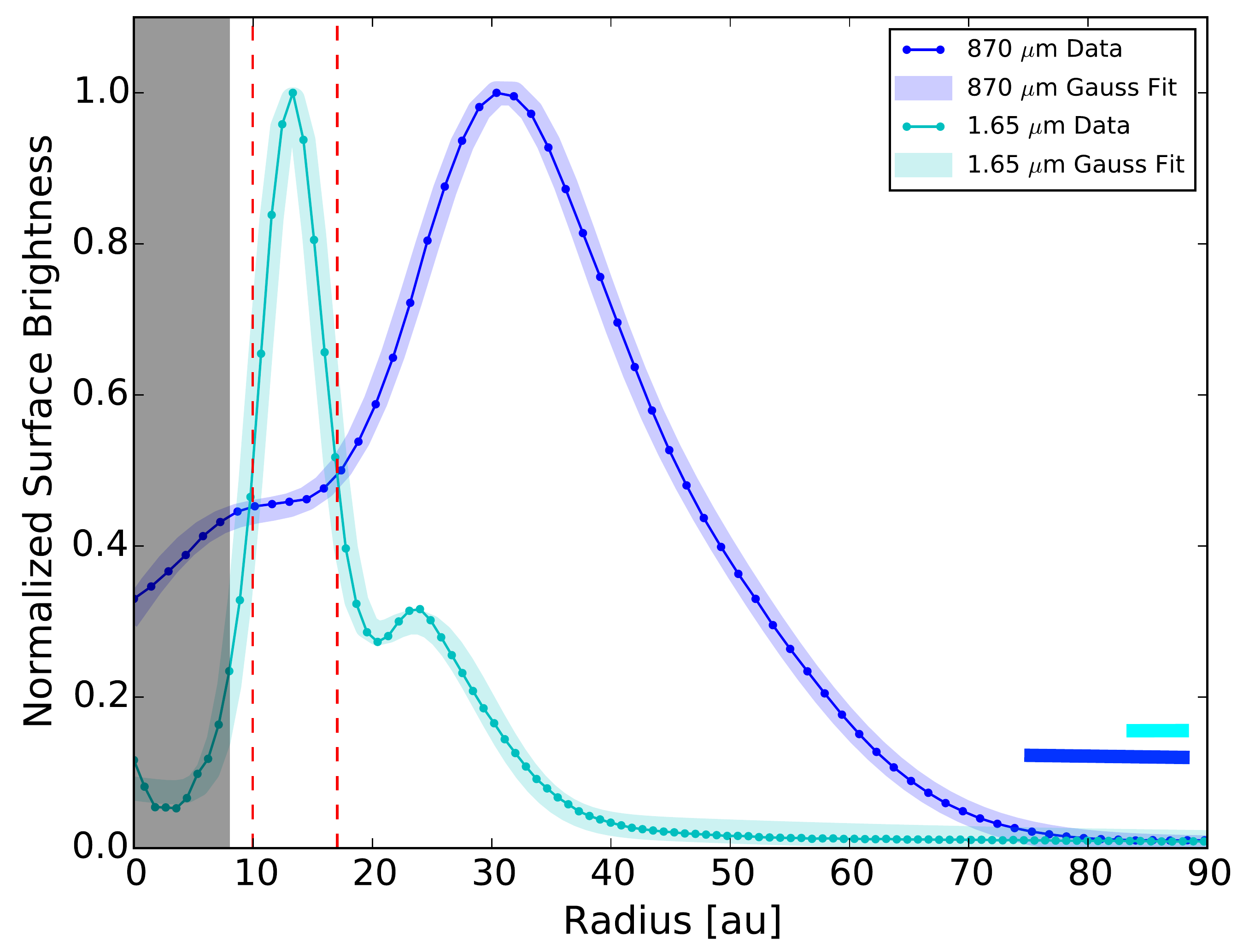}
        }%
    \end{center}
    \caption{  
    Comparison of the surface brightness profiles of the deprojected
    disk in the H-band and 870 $\micron$ continuum image (Figure
    \ref{Fig:ALMASPHERE002}) using a Briggs parameter of $-$2.0, with
    color coding and shading as in Figure
    \ref{Fig:Radialprofile}. Red vertical dashed lines indicate the
    FWHM of the inner ring imaged by SPHERE at H band. The blue and cyan bars in the lower right-hand corner indicate the spatial resolution at wavelength regime, i.e., $\sim$ 20 au for the ALMA data and $\sim$ 4 au for the SPHERE data, respectively.}

   \label{Fig:RadialProfile002}
\end{figure}

This higher-resolution ALMA 870 $\micron$ continuum image provides an additional point of comparison with
the SPHERE near-IR imaging. Figure \ref{Fig:RadialProfile002} shows
the normalized azimuthally averaged radial intensity profile extracted
from the $R = -2.0$ ALMA continuum image
overlaid on the SPHERE H-band radial profile. Following the analysis described in Section
\ref{Sec:RadialProfile}, we fitted Gaussian functions to the surface brightness
profile of the image obtained using $R = -2.0$. 
Perhaps not surprisingly, the resulting FWHM of the dominant Gaussian, 28.20 $\pm$ 0.04
au, is significantly smaller than that obtained using $R = -0.5$
($\sim$37 au; Table \ref{Table:Radialprofile}). More significantly,
after increasing the resolution of the continuum image obtained from
the ALMA data, the observed ring width ($\sim$28 au) now more
closely resembles the width of the outer ring obtained from the
synthetic images ($\sim$24 au; Table \ref{Table:Radialprofile}).

Interestingly, the radial profile obtained from the $R = -2.0$ ALMA continuum image shows a distinct shoulder between $\sim$
10 and 17 au, which closely corresponds to the position and width of the inner ring
traced by the NIR data (Fig.~\ref{Fig:RadialProfile002}). These spatially coincident features indicate that the inner ring includes a significant mass of mm-sized grains, in addition to the submicron-sized grains traced in scattered light.

\section{MODELING V4046 Sgr} 
\label{Sec:Modelling}

Our millimeter continuum image together with the scattered light image reveal a complex structure in the disk: an inner cavity surrounded by an inner ring, a dust gap, and an outer ring (Figures \ref{Fig:SPHEREALMA} and \ref{Fig:RGB}). The millimeter continuum image reveals a cavity and a bright outer ring with peak intensity at 32.30 $\pm$ 0.14 au, while the scattered light image displays an inner cavity with a radius of $\lesssim$ 9 au, an inner ring located at 14.10 $\pm$ 0.01 au, a gap at $\sim$ 20 au, and an outer ring coinciding with the inner edge of the mm continuum ring (Table \ref{Table:Radialprofile}). In the following, we explore the possibility that these features might be signposts of ongoing planet formation. Specifically, we investigate whether the gap at $\sim$20 au detected in the SPHERE scattered-light imaging and the ring/cavity structure imaged in the mm continuum by ALMA can be modeled in terms of interactions between the V4046 Sgr disk and a single planet. We note that the inner cavity in scattered light (and the associated inner ring at 14 au) is not included in our modeling. We only focus on the gap at $\sim$20 au, and investigate the hypothesis that it may be produced by a single planet.

\subsection{Nominal Model}

 The models we employ are the results of the 2D two-fluid (gas +
 particle) hydrodynamic calculations of planet-disk interactions
 combined with 3D Monte Carlo radiative transfer (MCRT) simulations
 presented in \citet{Dong2015}. In order to calculate the surface density distribution of the gas and dust, 2D (radial and azimuthal) two-fluid simulations were obtained using the FARGO code \citep{Masset2000}. The distributions of small and large grains are treated independently to generate the NIR and millimeter images. The model dust disk is primarily composed of small dust particles ($\sim$ 0.02 to $<$ 1 $\micron$) containing silicate, graphite, and amorphous carbon grains \citep{Kim1994}, with an additional component of large dust particles ($\sim$ 0.9 to $\sim$ 1 mm) containing 2/3 silicate (density 3.3 g cm$^{-3}$) and 1/3 graphite (density 2.3 g cm$^{-3}$). The surface brightness is calculated via the anisotropic scattering phase function developed by \citet{Henyey1941}, while the radiative equilibrium temperature stratification is calculated following the radiative equilibrium algorithm presented in \citet{Lucy1999}. To compute synthetic H-band (1.65 $\micron$)  scattered light and 870 $\micron$ continuum images, the radiative transfer package HOCHUNK3D is utlized \citep{Whitney2013}. More details of the disk setup and optical properties can be found in \citet{Dong2012} and \citet{Dong2015}.

\subsection{Constraints on Putative Planet Mass}
\label{Sec:Planetmass}

Since the width and depth of a gap that can be opened by a planet
depend most sensitively on planet mass, if disk viscosity and scale
height are fixed \citep[e.g.][]{Dong2017}, the characteristic surface
density profiles imposed on the disk for an ad hoc planet serve as a
guide to narrow the range of potential planet masses. Such an approach
was used to constrain the masses of putative planets forming in the
disks orbiting TW Hya and LkHa 330 by \citet{Rapson2015b} and
\citet{Uyama2018}, respectively. In the present case, we used the
library of fiducial hydrodynamical models presented in
\citet{Dong2015} to explore a range of ad hoc planet masses, examining whether their surface brightness profiles accurately reproduce the parameters measured for the scattered light (SPHERE) and thermal emission (ALMA) images (Table \ref{Table:Radialprofile}). Based on these tests, we established that the NIR  and mm-observations set the likely lower and upper planet mass bounds, respectively. We call such bounds Model 1 and Model 2.

In these models, and all other models we explored, the radial location
of the forming planet is fixed in a circular orbit at the location of the gap, i.e., 20 au. Because
the separation of the two stellar components is only 0.041 au, the
central source can be considered a single star of mass 1.75
M$_{\odot}$  \citep{Rosenfeld2012}.
We adopt the disk inclination of 32$^{\rm o}$ deduced from modeling
the continuum emission (Section \ref{Sec:IncPA}). To compare the
simulated images to the observations, we convolve the simulated H-band
images with a Gaussian point-spread function to achieve 0.04''
resolution ($\sim$ 3.0 au at 72.4 pc), while the simulated ALMA images
were produced by using \textit{simobserve} and \textit{simanalyze}
tasks in CASA to generate images convolved with the output clean beam
(0.29''$\times$0.17'' beam, or 20 au$\times$12 au at 72.4 pc). The
resulting synthetic images for Models 1 and 2 are shown in the middle and bottom panels
of Figure \ref{Fig:SPHEREALMA}, respectively. The
corresponding radial profiles obtained from the deprojected synthetic
images (extracted just as for the observations; see Section
\ref{Sec:RadialProfile}) are presented in the left (Model 1) and right
(Model 2) panels of Figure \ref{Fig:Radialprofile}, overlaid on the
observed radial profiles.

\paragraph{Model 1, 0.3 M$_{\rm Jup}$} A first inspection of the left
panel of Figure \ref{Fig:Radialprofile} shows that, overall, the surface
brightness profiles of the model disk structures resulting from
interactions with a 0.3 M$_{\rm Jup}$ planet --- i.e., the bright rings
and cavity in the model --- are comparable to the structures
detected by SPHERE (scattered light) and ALMA (thermal
emission). In particular, the gaps at $\sim$20 au have
similarly shallow depths in the observed and synthetic
scattered-light radial profiles, and the model well reproduces the
radial position of the outer scattered-light ring. As stated before, the inner cavity and the 14 au ring in scattered light is beyond the scope of our modeling effort here.

On the other hand, the 870 $\micron$ continuum
emission model generates a ring peaking at $\sim$27 au with a
FWHM of $\sim$24 au, which differ significantly from the
corresponding parameters characterizing the ALMA data (peak radius
$\sim$32 au, FWHM $\sim$37 au). Given that the emission cavity and ring diameter imaged in thermal emission at mm wavelengths scale with planet mass \citep{Dong2015}, these discrepancies between the observed and Model 1 ALMA images indicate that the putative planet mass is somewhat larger than 0.3 M$_{\rm Jup}$.

\paragraph{Model 2, 1.5 M$_{\rm Jup}$} 

In Model 2, the dust rings and gap in the disk have been displaced
radially, and as a result, the radii of the model and observed peak
intensities of mm-wave thermal emission in the outer ring of the disk
are in excellent agreement (Table \ref{Table:Radialprofile}). 

In this model, as in Model 1, we find a
width of $\sim$24 au for the mm-wave emission, meaning that although
the larger planet mass has resulted in a larger inner cavity, the
population of large grains is nevertheless still confined to a narrow
ring. However, in the case of the synthetic scattered light image, the
model gap at 20 au is wider and deeper than observed. 
 
This wider and deeper gap reflects the fact that, in Model 2, the planet is massive enough to open a gap whose surface density is reduced by several orders of magnitude relative to that of the local disk.

\section{Discussion}
\label{Sec:Discussion}

The semi-quantitative agreement between observations and models is encouraging (Figure
\ref{Fig:Radialprofile} and Table \ref{Table:Radialprofile}); these comparisons suggest that the gap around 20 au in the system may be opened by a Jovian planet, and the two models may bracket the mass of this putative planet. We emphasize that the models are not fine tuned to fit the observations; instead, they are generic models with parameterized profiles, and only the mass of the planet is varied in our limited explorations. Additional fundamental disk parameters, such as the viscosity, the scale height, and the overall gas surface density are all known to affect the gap profile in the gas and in mm-sized grains \citep[e.g.][]{Fung2014}; we have not explored the effects of varying these parameters in our modeling. In the following, we discuss specific aspects of (and some caveats to) the observational and modeling results described in the preceding sections, and their interpretation.

\subsection{A Planetary Companion}

In the MCRT images, the continuum millimeter synthetic images clearly
reveal a cavity and a bright outer ring with peak intensities at
$\sim$27 au and $\sim$32 au for 0.3 M$_{\rm Jup}$ and 1.5 M$_{\rm Jup}$, respectively. This reflects the fact that, in our modeling, the peak intensities in the millimeter are at larger radii as planet mass increases. The difference in the location of the peak intensities is not due to a radiative transfer effect; instead, it can be attributed to the millimeter emission that linearly scales with the surface density of the large grains ($\sim$1 mm). With our planet fixed in its orbit (i.e., 20 au) and with no variation in disk viscosity and scale height, the width of the cavity at millimeter wavelengths is determined mainly by planet mass, because the large grains are being trapped and piled up due to a gas pressure bump induced by the planetary companion \citep{Pinilla2012}. 

Similarly, in the simulated H-band images, the planet carves out a gap whose width and depth increases with planet mass (Figures \ref{Fig:SPHEREALMA} and \ref{Fig:Radialprofile}). Poor filtering of drifting dust explains the presence of micron-sized particles within and along the gap outer edge traced by the SPHERE data (Figures \ref{Fig:SPHEREALMA} and \ref{Fig:Radialprofile}). This is because the pressure maximum induced by a planet of only $\sim$ 1 M$_{\rm Jup}$ located at $\sim$ 20 au cannot efficiently filter out micron-sized particles  (1-10 $\micron$) at the outer edge of the planet gap. Such pressure peaks can slow down, but cannot completely stop the inward motion of the micron-sized particles \citep{Pinilla2016}.

Note that these simulations focus on reproducing observational signatures of a gap/cavity opened by a single planet fixed at 20 au and do not address possible signposts of multiple planets carving multiple gaps. It is also worth mentioning that the number of forming planets in the disk does not determine the number of gaps and instead, might be determined by the disk's viscosity \citep[e.g][]{Bae2018}. We caution that in these simulations, viscosity and scale height  (small and big grains) are taken to be constant. The choice of the viscous parameter, together with planet mass, determines the depth of the gap; specifically, decreasing the viscosity deepens the gap \citep{Dong2017}. We leave explorations of the dependence of the inferred putative V4046 Sgr circumbinary planet mass on disk viscosity and other disk properties (e.g., vertical dust settling, radial gradients in dust grain composition) to future work.

\subsection{Binarity, Disk Dispersal, and Disk Structure}

\subsubsection{Disk Lifetime}

We estimate that the ring imaged in mm continuum emission by ALMA comprises at least $\sim$60 M$_{\oplus}$ of dust  ( 3.1.1). Given the total (gas$+$dust) mass inferred for the circumbinary disk \citep[0.094 $M_\odot$;][]{Rosenfeld2013}, the disk most likely also includes a significant mass in smaller and larger dust particles that are undetectable in those data. In terms of this large disk gas mass, as well as its large molecular (CO) disk radius \citep[$\sim$300 au;][]{Kastner2018}, the V4046 Sgr disk stands as unique among members of the $\sim$20 Myr-old $\beta$ Pic Moving Group \citep{Riviere2014}. Indeed, among the four actively accreting, roughly solar-mass (K-type) star systems within $\sim$100 pc of the Sun that host long-lived yet gas-rich (protoplanetary) disks --- the other three being TW Hya, T Cha, and MP Mus --- V4046 Sgr hosts the most massive disk, even though it is by far the oldest of these four systems \citep{Sacco2014}. The V4046 Sgr disk is also the only {\it circumbinary} disk among these four nearby, long-lived protoplanetary disk systems.

Given this context, it is intriguing that the age of the V4046 Sgr system is
similar to the dust clearing time in a circumbinary disk predicted by
\citet{Alexander2012}. The simulations of the effects of binarity on
disk photoevaporation in that work showed that the tidal torque
effects generated by a binary star system may increase the lifetime of a circumbinary
disk by a factor of $\sim$ 3 for binary orbital separations significantly
smaller than the critical radius for disk photoevaporation.

A longer-lived disk could be a natural
consequence of such tidal torques, in the case of V4046 Sgr. Also, \citet{Kastner2011} have
speculated that the lifetime of the V4046 Sgr circumbinary
disk might have been extended by past interactions with an (M-type) tertiary
component that is now found at a separation of $\sim$12.5 kau from the
central close binary system. 

While the mechanism responsible for the apparent extended
circumbinary disk lifetime is still
unclear, it seems plausible that the close binary nature of V4046 Sgr
may have increased the time available for the formation of a planet,
thereby ultimately determining its properties. Planet formation relies
on the availability of time and material to establish the formation
of the first pebbles and planetesimals, but for circumbinary disks,
these initial stages might be postponed. This scenario could explain
the possible presence of a forming planet in a $\sim$ 20 Myr old
multiple system. Additional millimeter observations and detailed modeling are necessary to constrain disk dispersion and planet formation timescales in circumbinary disks, given that $\sim$ 40$\%$ of the identified population of disks that are in the process of dispersion (i.e., transitional disks) are found in close binary systems \citep{RuizRodriguez2016}.

\subsubsection{Binary - Disk Alignment}
\label{Sec:Aligment}

Tidal interactions at the inner edge of a circumbinary disk should force alignment of
the disk plane and binary orbital plane on timescales shorter than that of disk dispersal \citep{Bate2000, Foucart2013}, unless the central binary is eccentric and there exist large initial binary-disk misalignments \citep[e.g.][]{Zanazzi2018}. In the case of the (low-eccentricity) V4046 Sgr binary, such coplanarity had been previously inferred on the basis of
stringent constraints on the gas disk and binary inclinations obtained from CO kinematics and
optical spectroscopic
measurements, respectively \citep{Rosenfeld2012}. The inclination we
have inferred for the 870 $\mu$m continuum emission ring,
32.42$^{\circ}$$\pm$0.07$^{\circ}$ (Sec. \ref{Sec:IncPA}), now provides
additional evidence of coplanarity among the dust
disk, molecular gas disk, and central binary orbit. The timescale with which this
alignment was reached is $<$23 Myr (the age of V4046 Sgr), i.e., well before the
complete dispersion of the disk gas and, evidently, well before
the initiation of the formation of the 
putative Jovian planet whose mass we have constrained here. The V4046 Sgr system
hence stands as very strong evidence that the orbits of Jovian planets that have been spawned within circumbinary disks should be well aligned with the orbital planes of their host binary stars.

Finally, we note that, according to tidal truncation theory
\citep{Artymowicz1994}, a close binary star with a nearly circular
orbit may create a tidally-induced inner cavity in the protoplanetary
disk at $\sim$ 2.2$\textit{a}$ (where $\textit{a}$ is semi-major
axis). However, for V4046 Sgr, this implies a clear inner region of
only $<$ 0.1 au, two orders of magnitude smaller than the inner cavity
radius inferred from NIR observations.

\section{Summary}
\label{Sec:Summary}

We have presented 0.3$^{''}$ (20 au) resolution 870 $\micron$  ALMA
Band-7 observations of the circumbinary disk orbiting the 
V4046 Sgr close binary system. We re-analyze archival polarimetric SPHERE/IRDIS
\citep{Avenhaus2018} data, as a point of comparison with these new ALMA
observations. The combination of our new
millimeter continuum image and the scattered-light imaging
reveals the complex structure of V4046 Sgr's circumbinary dust disk.

The millimeter continuum
image reveal a cavity and a bright outer ring with peak intensity at
$\sim$32 au and extent of $\sim$90 au, while the scattered light image displays an inner cavity
with a radius of $\lesssim$9 au, an inner ring at $\sim$14
au, and an outer ring whose peak position ($\sim$25 au) corresponds
to the inner edge of the bright millimeter continuum emission ring. 

From a fit of the ALMA visibility data to a simple surface brightness model, we find
an inclination of 32.42$^{\circ}$ $\pm$ 0.07 for the bright 870
$\micron$ continuum ring, demonstrating that the large-grain dust
ring, larger-scale circumbinary gas disk, and central binary system are all in close alignment. 

From our 870 $\micron$ continuum flux density measurement of 880 $\pm$
40 mJy, we infer a lower limit on dust mass of $\sim$ 60.0
M$_{\oplus}$ ($\sim$ 0.2 M$_{\rm Jup}$). With the new ALMA observations, we also tentatively detected an inner ring located within $\sim$20 au of the central binary (Figures \ref{Fig:ALMASPHERE002} and \ref{Fig:RadialProfile002}).

We suggest that the complex radial distribution of dust in the V4046 Sgr circumbinary disk may be produced by an embedded planet. To this end, we
compared the combined ALMA+SPHERE observations with synthetic images
generated from generic planet-disk interaction models published in \citet{Dong2015}. The comparison shows that a planet-disk interaction model involving a 0.3 M$_{\rm Jup}$ planet at 20 au can reproduce the depth of the gap, as well as the peak and location of the second ring in the scattered light data, but somewhat underestimates the radius of the ring in the ALMA data. On the other hand, a model invoking a 1.5 M$_{\rm Jup}$ planet at the same location reproduces the location of the ring in the ALMA data, but greatly overestimates the width and depth of the gap in the scattered light data. These results hence suggest that the mass of the putative circumbinary planet orbiting V4046 Sgr at $\sim$20 au should lie within the range 0.3--1.5 M$_{\rm Jup}$  (Figures. \ref{Fig:SPHEREALMA} and \ref{Fig:Radialprofile}). We note however, that the models we have employed here were not specifically designed to reproduce the V4046 Sgr disk. Furthermore, these models do not account for the inner ring and cavity in scattered light, which may be produced by additional planets inside 15 au.

We encourage additional detailed modeling of the V4046 Sgr disk, to further test this hypothesis. Specifically, the mass
of the putative circumbinary planet orbiting V4046 Sgr can be further
constrained by exploring the dependence of the inferred planet mass on
disk properties such as viscosity, vertical dust settling, and dust
grain composition. Higher resolution ALMA continuum
imaging is also required to confirm the presence of the inner 15 au radius ring that is marginally detected and resolved in the image presented in Figure \ref{Fig:ALMASPHERE002} . If confirmed, this inner ring may indicate the presence of a second circumbinary planet in the system, interior to 15 au. Regardless, the analysis presented here  illustrates the unique insights into the formation and early evolution of circumbinary
planets that can be obtained via studies of disks orbiting
V4046 Sgr and other young binary systems.

We are grateful to an anonymous referee for constructive suggestions that improved our paper. This paper makes use of data from ALMA program
ADS/JAO.ALMA No. 2016.1.00315.S. ALMA is a partnership of
ESO (representing its member states), NSF (USA) and NINS (Japan),
together with NRC (Canada), NSC and ASIAA (Taiwan), and
KASI (Republic of Korea), in cooperation with the Republic of
Chile. The Joint ALMA Observatory is operated by ESO, AUI/NRAO
and NAOJ. The National Radio Astronomy Observatory is a facility
of the National Science Foundation operated under cooperative
agreement by Associated Universities, Inc. D.A.R. and J.H.K
acknowledge support from NASA Exoplanets Program grant 
NNX16AB43G to Rochester Institute of Technology.

\newpage

\bibliographystyle{aasjournal}
\bibliography{biblio}

\end{document}